# Study of the neutron dose equivalent at the HIRFL deep tumor therapy terminal

Jun-Kui Xu (徐俊奎)[1,2;1)]  You-Wu Su (苏有武)[1]  Wu-Yuan Li (李武元)[1]  Wei-Wei Yan (严维伟)[1]

Zong-Qiang Li (李宗强)[1]  Wang Mao (毛旺)[1]  Cheng-Guo Pang (庞成果)[2]  Chong Xu (徐冲)[1]

[1]Institute of Modern Physics, Chinese Academy of Sciences, Lanzhou 730000, China

[2]School of Nuclear Science and Technology, Lanzhou University, Lanzhou 730000, China

**Abstract:** The secondary neutron fields at the deep tumor therapy terminal at HIRFL (Heavy Ion Research Facility in Lanzhou) were investigated. The distributions of neutron ambient dose equivalent were measured with a FHT762 Wendi-II neutron ambient dose equivalent meter as the $^{12}$C ions with energies of 165, 207, 270, and 350 MeV/u bombarded on thick tissue-like targets. The thickness of targets used in experiments is larger than the range of the carbon ions. The neutron spectra and dose equivalent is simulated by using FLUKA code, and the results agree well with the experimental data. The experiment results showed that the neutron dose produced by fragmentation reactions in tissue can be neglected in carbon-ion therapy, even considering their enhanced biological effectiveness. These results are also valuable for radiation protection, especially in the shielding design of high energy heavy ion medical machines.

## 1 Introduction

Carbon ion is the most common particle in heavy ion tumor therapy due to its physical and biological properties. The so called "Bragg peak" of heavy ions makes it very suitable for radiotherapy. However, secondary particles caused by heavy ion reactions must be taken into account in the shielding design of heavy ion medical machines and personal dose assessment. Neutrons are the most abundant products in all secondary particles and very important in safety evaluation for radiation protection. Furthermore, the neutrons can affect a large area due to its strong penetrating power, that is to say, it can influence the whole body of the patient including tumor and healthy tissue.

In recent years, there are many studies on the secondary neutrons induced by mid-energy heavy ions and proton bombarding on thick targets. For example, Sunil *et al*. measured the neutron dose equivalent in low energy (about 7 MeV/u) heavy ion interactions with thick targets [1, 2]. In 2012, Nandy *et al*. measured the neutron dose equivalent distribution in carbon ions induced reactions with Ti and Ag targets using proton recoil scintillator [3]. In view of radiation protection of patients, Haluk Yucel *et al*. measured the neutron dose equivalent at the 18-MV medical linac [4] in 2015. Howell *et al*. measured neutron dose equivalents in the craniospinal irradiation by a passively scattered proton system [5]. Saeed *et al*. calculated the ambient dose equivalent of fast neutrons by using the elemental composition of human body and the conversion coefficient ICRP116 [6]. T. Kurosawa *et al*. performed a systematic experimental study on neutron yields, spectra and angular distributions of various ions on thick targets with TOF method at HIMAC (Heavy Ion Medical Accelerator in Chiba), and compared their experiments with the HIC code calculation results [7, 8].

In our previous work, the secondary neutrons induced by heavy ions were also studied. G. Li *et al*. measured neutron yields by 50-100 MeV/u heavy ions hitting on thick targets using activation method at HIRFL IMP [9]. Neutron dose equivalent distributions have been measured at the superficial tumor treatment terminal of HIRFL. In that experiment, carbon ions were accelerated to 100 MeV/u by separated-sector cyclotron (SSC) of HIRFL, and then delivered vertically down to a basement where the superficial tumor treatment terminal located [10]. These studies are useful for radiation protection such as neutron shielding design, and also have significance in tumor therapy.

The maximum energy of $^{12}$C ions is 100 MeV/u when the ions were extracted from SSC, which is not high enough for deep tumor therapy. However, HIRFL-CSR could be used to accelerate the carbon ions to higher energies (maximum energy is 900 MeV/u) for studying the deep tumor therapy. Fig. 1 shows the layout of the HIRFL-CSR and the location of the deep tumor therapy terminal. In the present work, the neutron dose was measured at this deep tumor therapy terminal as carbon ions with specific energy of 165, 207, 270 and 350 MeV/u bombarding on thick tissue-like targets. Then we calculated neutron dose distribution and energy spectra with FLUKA code. It is found that the measured neutron dose results have a good agreement with the calculated results.

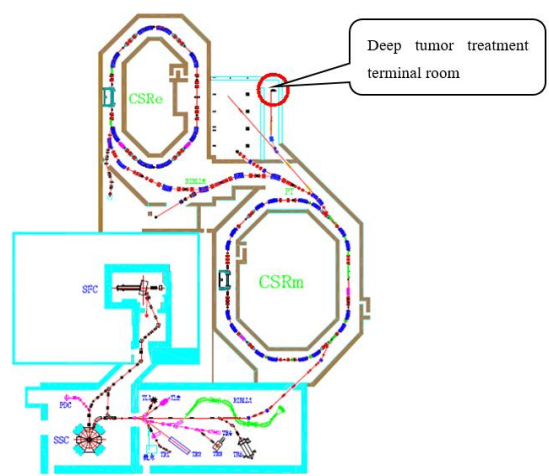

Fig.1 (Calor online) Schematic diagram of the HIRFL-CSR.

## 2 Experimental apparatus

The experiment was carried out at the deep tumor therapy terminal of HIRFL-CSR. $^{12}$C ions were accelerated to 80 MeV/u by SSC then injected into CSRm for accelerating to energies needed. In this work, the carbon ion energies were chosen as 165, 207, 270 and 350 MeV/u. In order to measure the ion counts, the extracted $^{12}$C ions firstly penetrated a parallel plate ionization chamber detector, in which the nitrogen gas with an atmosphere pressure is filled, and in this work the detector was used for counting the primary particle numbers. Then the ions were delivered to the deep tumor therapy terminal. A tissue-like target which is thick enough to stop the ions was placed on the treatment bed. Table 1 lists the material composition of the target. The target is the superposition of several thin slices, whose cross section is 30 cm × 30 cm. Table 2 shows the thickness of the tissue-like target used for different energy carbon ions incidence.

For monitoring the produced neutrons a modified A-B rem-meter detector [11] was placed at the same height as the target used (1.2 meters high above the floor) in this experiment. Here, we called it FHT672 Wendi II neutron ambient dose equivalent meter, which uses a large volume He-3 tube and could measure neutrons with energies ranging from thermal to 5 GeV. Calculations show that adding heavy metal material layer in rem-meter can increase the neutron response in high energies while keeping the response curve unchanged in the lower energy region [12-14]. It should be noted that there is an uncertainty in measuring the high energy neutron dose because of the lack of the mono-energetic neutron sources for calibrating the detector. However, if the neutron spectra could be simulated it is possible to obtain the neutron dose by correcting the measured results with a standard A-B rem-meter. Fig. 2 shows the experimental arrangement and the Wendi II neutron ambient dose equivalent meter.

Table 1. Material composition of the tissue-like target, the density is 1.043 ±0.005 g/cm$^3$.

| Component | H | C | N | O | Cl | Ca |
|---|---|---|---|---|---|---|
| Proportion (%) | 8.1 | 67.2 | 2.4 | 19.9 | 0.1 | 2.3 |

Table 2. The thickness of the tissue-like targets used for different carbon ion energies.

| Ion energy (MeV/u) | 165 | 207 | 270 | 350 |
|---|---|---|---|---|
| Target thickness (cm) | 8 | 12 | 18 | 26 |

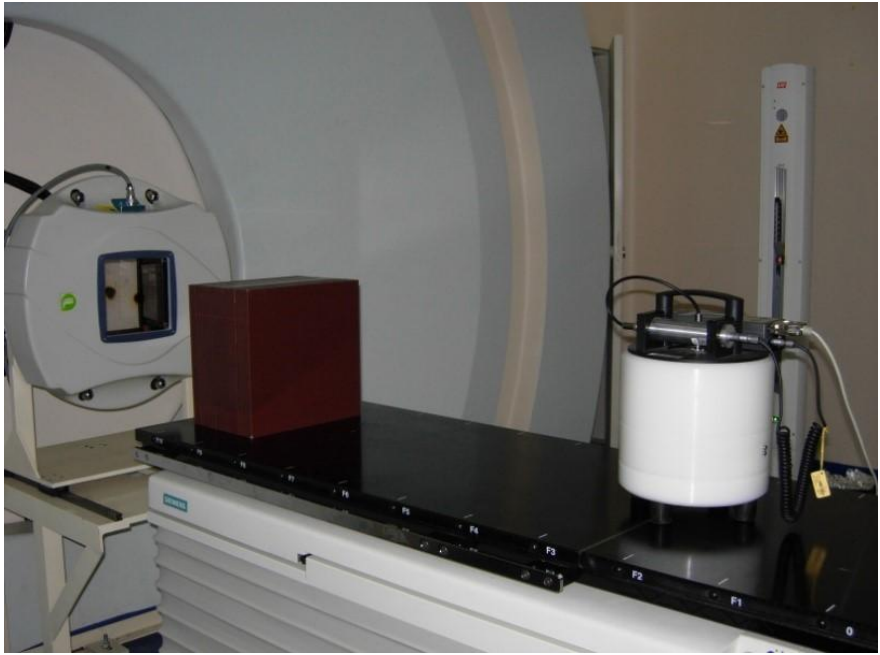

Fig. 2 (Calor online)The experimental arrangement and the Wendi II neutron ambient dose equivalent meter used.

In the measurements, as the parallel plate ionization chamber detector's counts are accumulated to the number of $4 \times 10^5$, the primary beam will be automatically cut off, and the FHT672 Wendi-II neutron ambient dose equivalent meter will stop recording the neutron dose equivalent. Then the Wendi-II detector is shifted to another observation position and the primary beam is extracted again. Using this method, neutron dose emitted from the target was measured at 0 Deg., 30 Deg. and 90 Deg. at different distances.

**3 Monte Carlo simulation**

FLUKA code [15, 16] was used to simulate the neutron dose equivalent distribution and energy spectra in the present work. In this code, the nucleus-nucleus interaction was described by the Dual Parton Model (DPM) [17] for ion energies over 5 GeV/u, the Relativistic Quantum Molecular Dynamics Model (RQMD) [18] for energies from 0.1 GeV/u to 5GeV/u and the Boltzmann Master Equation (BME) [19] theory for energies below 0.1 GeV/u, respectively.

The detailed description of FLUKA code can be seen in the previous paper [15]. In our simulation, parallel $^{12}$C beam was delivered to the tissue-like target which has the same size and composition as used in the experiment (table 1 and table 2). The size of the treatment room filled with dry air is 4 m $\times$ 5.6 m $\times$ 8 m. The simulation geometry only includes the shielding wall and the target, other accessories are neglected for a simplification. The cross section diameter of the mono-energetic beam is 5 mm with Gaussian distribution. In a sampling process one carbon ion is extracted from the simulation source. The tissue-like target is bombarded by carbon ions with energy of 165, 207, 270 and 350 MeV/u, here, the selected ion energies and the targets are consistent with the experimental parameters. In the process of simulation, the PHSICS card was used to select the specific class of transport problems, such as particle transport threshold, multiple scattering threshold, and heavy fragment transport activated and so on. The USRBDX card was used to obtain the neutron energy spectra, and set of detectors were placed at 1 meter far from the injection point at different angles as shown in Fig. 3 in the Monte Carlo simulation. The total histories in the simulation were from $10^9$ to $8 \times 10^9$ according to projectile energies. The relative error of the calculation was less than 5%. The USRBIN card was chosen to obtain the neutron dose equivalent [15]. FLUKA version 2011.2c was used in this work.

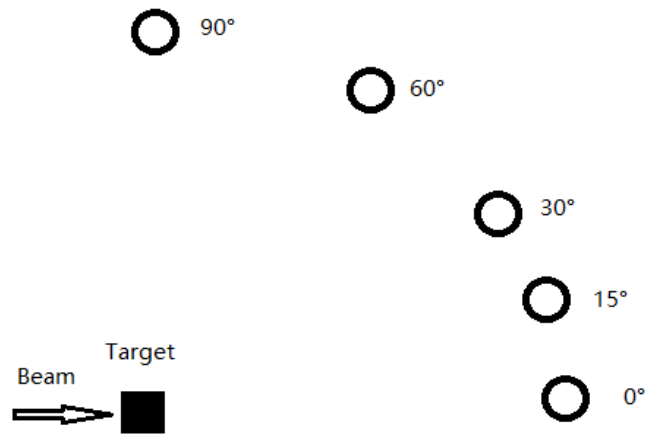

Fig. 3 The geometric arrangements for calculating the neutron energy spectra.

**4 Results and discussion**

Fig. 4 shows the measured neutron ambient dose equivalent at 0°, 30° and 90° with respect to the beam direction as a function of the distance from the target to the FHT672 Wendi-II detector. The solid square represent the neutron dose equivalent induced by carbon ions with energy of 350 MeV/u, the circle, up-triangle and down-triangle represents the dose by carbon ions with energies of 270 MeV/u, 207 MeV/u and 165 MeV/u, respectively. The parallel plate ionization chamber was used as a counter. For carbon ion energies from 165 MeV/u to 400 MeV/u, one count of the ionization chamber represent about 900 to 1300 carbon ions. So the chamber will approximately make a response as 1100 carbon ions with an error of ±18% penetrating it. According to the calibration test report, the linear error of the FHT672 is from -9% to 11% and the angle dependence error is about ±20%. Therefore, the error of the data measured by FHT672 Wendi-II neutron meter is from -22% to 23%. The total uncertainty of the dose as showed in Fig. 4 is less than ±30%.

As can be seen from Fig. 4, the measured results show the following characteristics: (1) At the same position in the deep tumor terminal room, the neutron dose equivalent increases with the increasing of the carbon ion energies. This is because the neutron yields produced by the fragmentation reaction increases as increasing the carbon ion energies. (2) With the same incident energies of the carbon ions, the neutron dose equivalent exponential decreases approximately with increasing the distance from the dose meter to the tissue-like target center. Moreover, the decrease shape becomes more approximately consistent in the larger angle. (3) It also can be seen that the neutron dose at $0^o$ is about 100 times larger than the dose at $90^o$ for the same incident energies. This can be explained that the neutron yields in small angle is larger than the large angel, and the cascade neutrons were the major part of the neutron yield, i.e., the neutron emission has a sharp forward peaking. In large angle, the measured neutrons are mainly from low energy evaporation mechanism.

In order to understand the cumulative amount of neutron dose equivalent in a simplified treatment process, we can make a qualitative estimate. The accumulated counts measured by the parallel plate ionization chamber in an ordinary carbon ion therapy with energy of 350 MeV/u are about $1 \times 10^7$. Therefore, the accumulate neutron dose equivalent at 20 cm far from the target center at $0^o$ is about 15 mSv. Based on this number the neutron absorbed dose is estimated as a few of mGy, which is a low level compared with the treatment dose in carbon ion therapy [16]. This indicates that the neutron dose produced by fragmentation reactions in tissue is small and even can be ignored, even though considering their enhanced biological effectiveness in radiotherapy.

The FLUKA simulated results of the neutron dose distribution are shown in Fig. 5. The one on the left of the top row represents the results induced by 165 MeV/u carbon ions, and the right one is by 207 MeV/u carbon ions. In the following row, the left one represents the distribution by 270 MeV/u carbon ions, and the right one is by 350 MeV/u. It can be seen that the simulated results exhibit the same tendency with the measured one, i.e., the neutron dose equivalent increases as the projectile energy increases, and the dose at $0^o$ are much higher than that at $90^o$. It should be noted that the neutron dose equivalent at $0^o$ near the shielding wall increases with the increase of the incident energies. This result could be explained by the (n, xn) spallation reaction happened as the high energy neutrons bombarding the concrete wall [17], and the neutron yield will increase as the carbon ion energies increase.

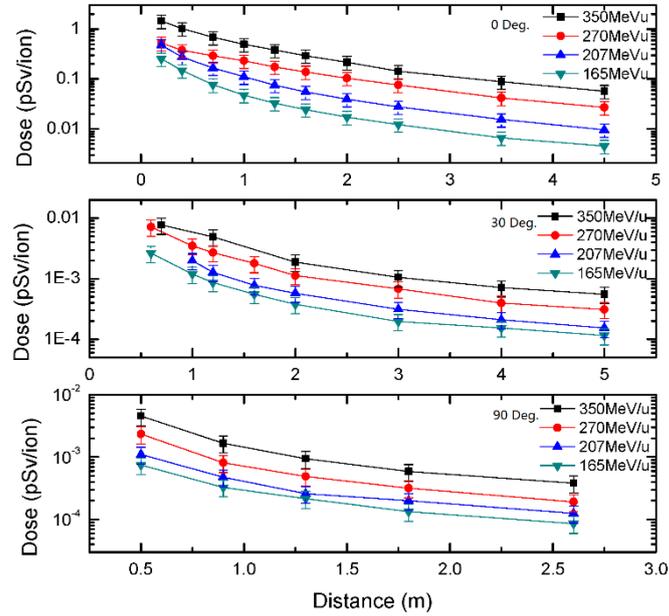

Fig.4 (Calor online) The measured neutron dose equivalent as a function of the distance from the center of the targets to the FHT672 Wendi-II neutron meter for observation angles of 0º, 30º and 90º, respectively. Square, circle, up-triangle and down-triangle represents the carbon ion with energies of 350, 270, 207, 165 MeV/u, respectively.

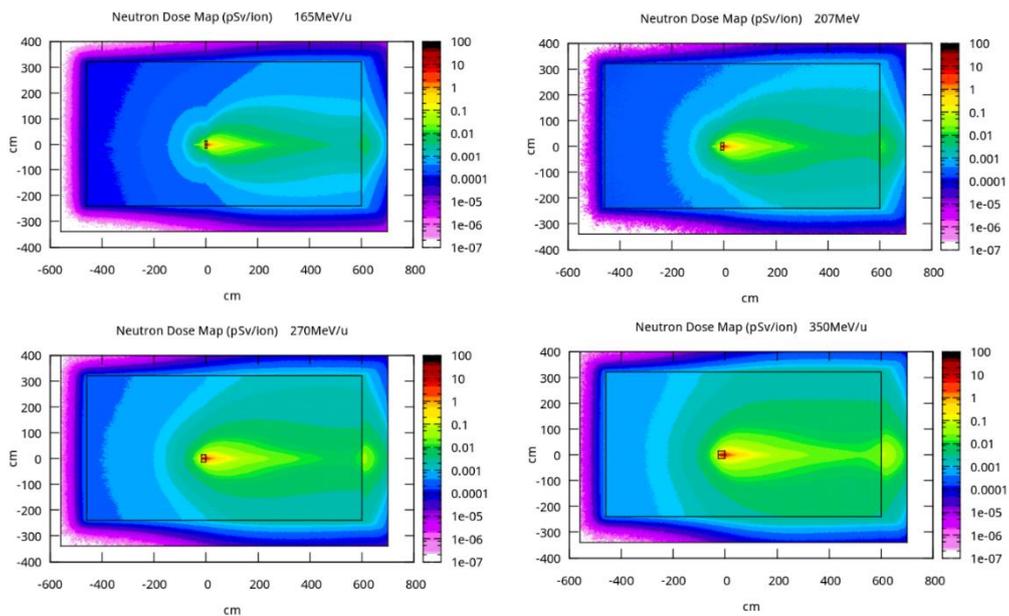

Fig. 5 The calculated neutron dose equivalent distributions induced by carbon ions with energies of 165 MeV/u, 207 MeV/u, 270 MeV/u and 350 MeV/u, respectively.

The neutron does equivalent at the 0º calculated with the FLUKA code is shown in Fig. 6 together with the experimental results. The solid symbols represent the measured data, and the lines represent the FLUKA simulations. The shapes of the simulated results are agreed well with the measured ones, but the absolute values are about 2.1 times of the experimental data. In Fig.6 the experimental data are shown in absolute value. The discrepancy could be from the following reasons. Firstly, the detector itself cannot achieve the ideal requirement, namely its energy response to neutrons is not absolutely agreed with the human response. The second reason is related to the uncertainty of the primary ion counts. Finally, the differences are maybe from the program code itself.

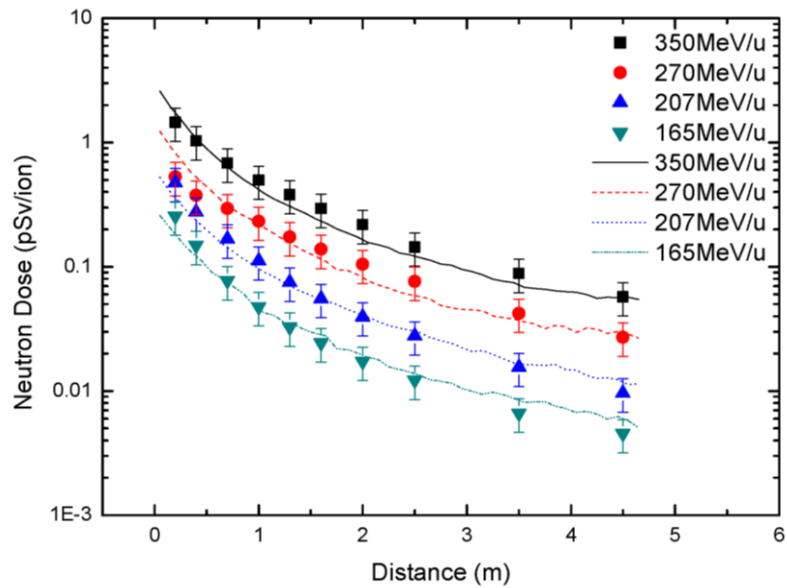

Fig. 6 (Calor online) The experimental and the calculated neutron dose equivalent at 0º for different incident energies. The symbol represents the experiments: square, circle, up-triangle and down-triangle represents the carbon ion with energies of 350, 270, 207, 165 MeV/u, respectively. The line represents the corresponding simulation.

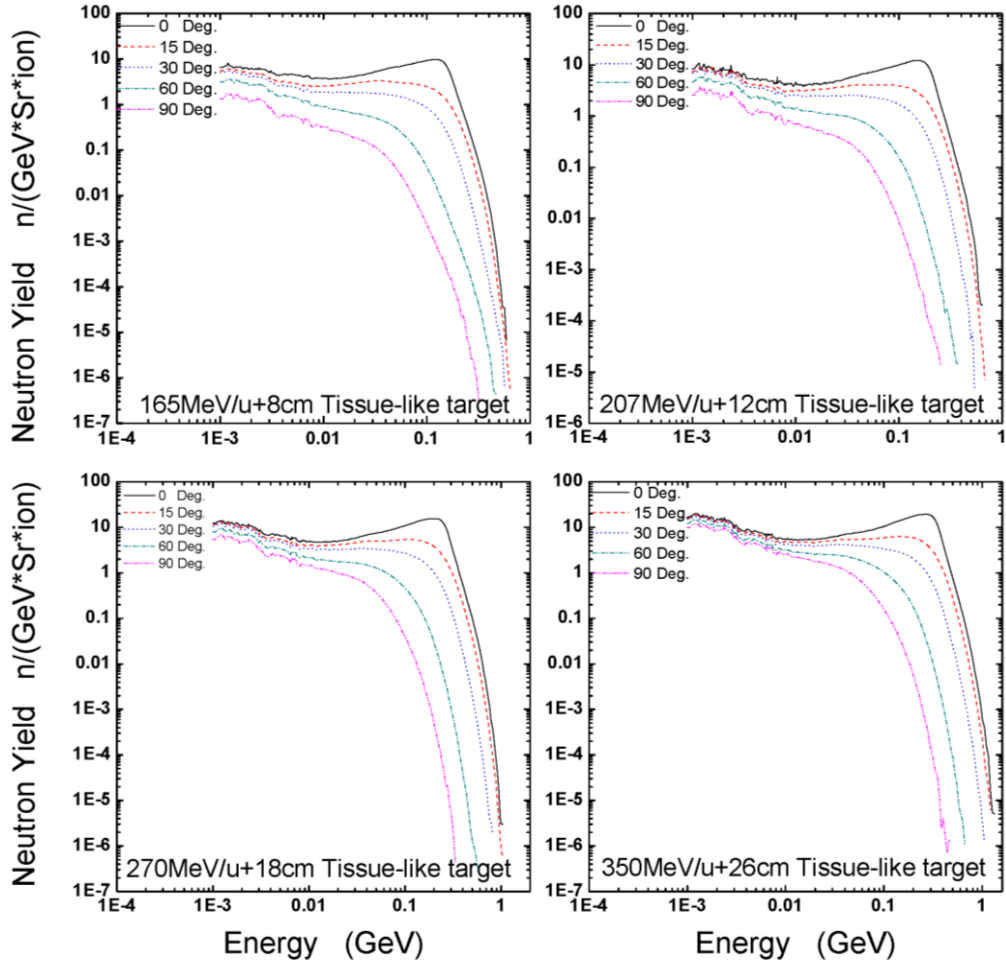

Fig. 7 (Calor online) The calculated neutron energy spectra of $^{12}$C ions bombarding on tissue-like target.

In order to further interpret the measured neutron dose equivalent, the neutron energy spectra were also calculated with the FLUKA code for $^{12}$C ions with energies of 165 MeV/u, 207 MeV/u, 270 MeV/u and 350 MeV/u, respectively, bombarding on thick tissue-like target. It should be noted that the neutron dose can be calculated from the neutron spectra by using the flux to dose conversion factors (H*(10)) of ICRP publication number 74[18]. The calculated results are shown in Fig. 7. As can be seen from the figure, there has a board peak at the high energy range in the forward direction, and the yield of high energy neutrons decreases swiftly as increasing the angels. This result means that the measured higher intensity neutrons at $0^o$ as mentioned above are high energy neutrons, which are produced in the cascade (or pre-equilibration) process. It also can be seen that the neutrons below 10 MeV are approximately independent on the energies of primary carbon ions and the ejection angles. This means that these neutrons are produced isotropically by equilibrium (or evaporation) mechanism.

# 5 Conclusions

In the present work, we measured the neutron dose distribution at the HIRFL deep tumor treatment terminal as 165, 207, 270 and 350 MeV/u 12C ions, respectively, bombarding on tissue-like targets. It is found that the neutron dose exponentially decreases approximately with the distance from the target to the tissue-like target, and the dose at 0º is about 100 times larger than that at 90º for same distances. The experimental fact indicates that the neutron dose equivalent produced by fragmentation reactions in tissue is small as compared with the treatment dose. We also calculated the neutron dose distribution by using the FLUKA code. The results were in agreement with the experimental data except for a constant factor. The secondary neutron energy spectra were also calculated. The measurements and the simulations are valuable for radiation protection, especially in the shielding design of high energy heavy ion medical machines and the personal dose assessment.